\documentclass[showpacs,reprint,preprintnumbers,amsmath,amssymb,superscriptaddress,longbibliography,nofootinbib]{revtex4-2}

\usepackage[utf8]{inputenc}
\usepackage{amsmath}
\usepackage{amssymb} 
\usepackage{enumerate}
\usepackage{multirow}
\usepackage{braket}
\usepackage[toc,page]{appendix}
\usepackage{url}
\usepackage{graphicx}
\usepackage[pdftex,bookmarks,linktocpage,pdfpagelabels,plainpages=false,hyperfigures,linkcolor=blue,citecolor=blue]{hyperref} 
\hypersetup{colorlinks=true}

\usepackage{mathrsfs,amssymb}  
\usepackage{cancel}
\usepackage[normalem]{ulem}
\usepackage{cleveref}
\usepackage{tikz}
\usetikzlibrary{decorations.markings}
\usetikzlibrary{positioning}
\usetikzlibrary{shapes}
\usetikzlibrary{calc}

\newcommand{\beq}{\begin{equation}}
\newcommand{\eeq}{\end{equation}}

\newcommand{\bea}{\begin{eqnarray}}
\newcommand{\eea}{\end{eqnarray}}

\newcommand{\QUBO}{{\rm QUBO}}

\begin{document}
\date{November 15, 2021}

\preprint{KIAS--P21050}

\title{Leveraging  Quantum Annealer to identify an Event-topology at High Energy Colliders} 
\author{Minho Kim}
\email{mhkim90@kias.re.kr}
\affiliation{Quantum Universe Center, KIAS, Seoul 02455, Korea}
\author{Pyungwon Ko}
\email{pko@kias.re.kr}
\affiliation{School of Physics, KIAS, Seoul 02455, Korea}
\author{Jae-hyeon Park}
\email{jhpark@kias.re.kr}
\affiliation{School of Physics, KIAS, Seoul 02455, Korea}
\author{Myeonghun Park}
\email{parc.seoultech@seoultech.ac.kr}
\affiliation{School of Physics, KIAS, Seoul 02455, Korea}
\affiliation{Institute of Convergence Fundamental Studies, Seoultech, Seoul, 01811, Korea}

\begin{abstract}
With increasing energy and luminosity available at the Large Hadron collider (LHC), we get a chance to take a pure bottom-up approach solely based on data.
This will extend the scope of our understanding about Nature without relying on theoretical prejudices.  
The required computing  resource, however, will increase exponentially with data size and complexities of events if one uses algorithms based on a classical computer. 
In this letter we propose a simple and well motivated method with a quantum annealer to identify an event-topology, a diagram to describe the history of particles produced at the LHC. 
We show that a computing complexity can be reduced significantly to the order of polynomials which enables us to decode the ``Big" data in a very clear and efficient way. 
Our method achieves significant improvements in finding a true event-topology, more than by a factor of two compared to a conventional method.
\end{abstract}

\maketitle
\noindent{\bf Introduction.} Understanding data has been always a milestone to build the theoretical understanding of our universe. 
When we have a strong theoretical motivation, we design an experiment to confirm it. 
The recent discovery of the Higgs particle is the perfect example as we have launched high energy collider programs to identify a mechanism behind electroweak symmetry breaking\,\cite{ATLAS:2012yve, CMS:2012qbp}. 
But when existing theories have not been supported by experiments,  unbiased observations on phenomena can shed light on a way to expand our theoretical framework. 
With this bottom-up approach, one can take Occams' razor. For example, the idea of dark matter emerged as a simple explanation for the observed anomaly in various galaxy rotation curves\footnote{One can find a good review about the history of dark matter\,\cite{Bertone:2016nfn}.}. 

The standard model of particle physics (SM) had been constructed and confirmed by various experiments, finally was proved by the Higgs discovery but its extension is required to accommodate 
new phenomena including dark matter.
As one of robust experiments, the LHC has probed various extensions of SM so far. 
The most interesting result  of the LHC is that it opens a new opportunity for the young generation by closing the window for the previously favorite models including weak scale supersymmetric models.
To cope with the current challenge of being without any theoretical guidelines, the LHC community starts to consider model-independent and data driven methods\,\cite{LHCNewPhysicsWorkingGroup:2011mji}.

Being independent of a theory requires huge computing power as one needs to check the whole possibilities hidden in big data. 
This is the very reason to adopt the art of modern computing, a machine learning (ML) to data analyses. 
The high energy physics (HEP) community has taken the most sophisticated ML algorithms and achieved significant progresses\footnote{Well maintained list of ML applications can be found in \cite{Feickert:2021ajf}.}.
Even with these successes, there would be the fundamental gap in applying ML to the problem of HEP;
(1) Current ML algorithms are optimized to extract features in commercial data.  (2) The size of data from the LHC exceeds the size of commercial one 
by the factor of $\mathcal{O}(10^2)$\,\cite{Girone:2315475}.
(3) The memory bottleneck in processing big data is rooted in the current computing architecture\,\cite{kumar2019efficient, ojika2020addressing}.

Unlike the current ``classical" computer, a quantum computer has the advantage to resolve above issues by the core concepts of quantum physics; superposition principle, entanglement and quantum tunneling\,\cite{Preskill:2012tg}. 
There have been pioneering works in applying quantum computing including\,\cite{Mott:2017xdb, Wei:2019rqy, Abel:2020ebj, Pires:2020urc, PhysRevLett.126.062001,Bauer:2021gup}.
In this letter, we would like to add our effort by proposing a simple but powerful algorithm to perform the LHC data analysis with a very minimum assumption by the help of a quantum computer.

If there is an anomaly in the LHC data, one needs to identify the structure of signals as we draw a Feynman diagram to identify an underlying signal process. 
An event-topology is a Feynman diagram-like drawing without the spin information. With this, we focus on a kinematic structure of how observed particles are grouped according to decays of intermediate particles some of which may be new states not present in the SM\@. Once we identify an event-topology, we can measure efficiently masses and spins of new particles involved in the process\,\cite{Barr:2011xt, Wang:2008sw}. 
As we will show later, identifying an event-topology in an unbiased way requires enormous computing power from combinatorics. We resolve this issue by converting the combinatorial problem into a quadratic unconstrained binary optimization (QUBO) problem.  The price for this is to handle with a notorious local minimum problem, the most significant issue in current ML algorithms\,\cite{li2017visualizing}. Here we make use of a quantum annealer thereby 
exploiting the quantum advantage to resolve this issue.

\begin{figure}[t]
\centering
\includegraphics[width=0.48\textwidth]{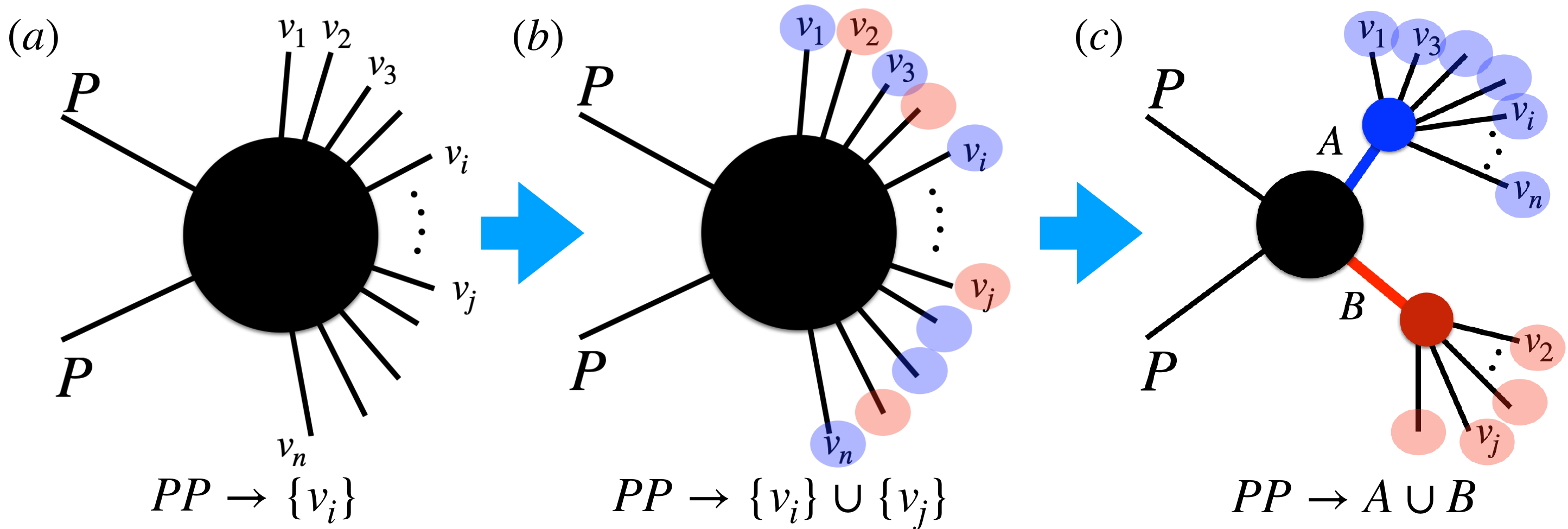}
\caption{(a) $n$-observed particles (b) Dividing $n$ particles into two groups for $2\to2$ process (c) Identified event-topology with $A$ and $B$.
\label{diagram}
}
\end{figure}

\noindent{\bf QUBO for Event-topology classification.} Our only assumption on abnormal events in collider data is that observed particles are produced through $2\to 2$ process. More specifically, two new particles $A$ and $B$ are produced and they decay into observed ones. Thus identifying an event-topology becomes a binary classification, whose computing complexity increases exponentially as $\mathcal{O}(2^n)$ with $n$ observed particles.  A schematic description is presented in Fig.\,\ref{diagram}.  As we have no further assumptions, we need to set a guiding rule to assign observed particles into decay products of either $A$ or $B$. 
Motivated by general ``energy minimum principle" in various fields of physics, one attempts to minimize the total invariant mass $(P_1 +P_2)^2$. 
But unlike the case of signals with missing energy which have been studied extensively in the literature, we will have a trivial partonic center of energy $\sqrt{\hat s}$ 
when all the final particles are visible without missing energy-momenta. 

The next trial we can take is to minimize a mass 
difference between $A$ and $B$.  With the four-momentum of $i$-th particle as $p_i$, momentums of $A$ and $B$ are;
 \beq
P_1 = \sum_i p_i\,x_i,~ P_2 = \sum_i p_i\,(1-x_i), 
\eeq
where
$p_i$ is the constituent of $A$ ($P_1$) if $x_i=1$ or $B$ ($P_2$) if $x_i=0$\,\cite{Wei:2019rqy}.
Unlike a jet clustering algorithm, we don't require any structure or a seed in clustering particles. 
By focusing on the kinematics, we minimize the following function $H$, the mass difference of $A$ and $B$;
\beq 
H=\left(P_1^2-P_2^2\right)^2
\label{eq:Hmin}
\eeq
for all possible combinations of $\{x_i\}$. 
The dimension of $H$, $[H] = M^4$ is chosen to address our problem as a QUBO problem with an Ising model form;
\beq
H_{\rm QUBO} = \sum_{ij} J_{ij} s_is_j + \sum_ih_is_i,
\label{eq:QUBO}
\eeq
where $\{s_i\}$ is spin set with only $\pm1$ values for spin  $\uparrow$ and $\downarrow$, and $J_{ij}$, $h_i$ are the coupling strength and biases, respectively.
We cast our minimization problem on $H$ into that on $H_\QUBO$ through a change of variables $x_i=(1+s_i)/2$ to express;
\bea
&&J_{ij} = \sum_{k\ell}P_{ik}P_{j\ell}, \\
&&h_i = 2\sum_j[\sum_{k\ell}(P_{ik}P_{j\ell} -P_{k\ell}P_{ij})] ,
\eea
with $P_{ij}=p_i \cdot p_j$. Our target function $H$ in Eq.\,(\ref{eq:Hmin}) is optimized to the case of $M_A = M_B$, which is the case of most conventional new physics searches at the LHC. Thus this function $H$ can be a starting point, but we need to generalize this function to handle situations including (1) various new physics scenarios with asymmetric production of $M_A \neq M_B$, and (2) off-shell effect from the decay width of unstable particles or smearing from a detector responses. We add an additional constraint term to deal with above issues;
\bea
H_\QUBO&\rightarrow& H_\QUBO + \lambda(P_1^2+P_2^2)\nonumber\\
&=&H_\QUBO + \lambda \sum_{ij} P_{ij} [s_i s_j + (1-s_i)(1-s_j)]\nonumber\\
&=&\sum_{ij} J_{ij}' s_is_j + \sum_ih_i'\,s_i,
\label{eq:QUBO_lambda}
\eea
with $J_{ij}' = J_{ij}+ 2\lambda P_{ij}$ and $h_i' = h_i - 2 \lambda \sum_j P_{ij}$.  Here we remove constant terms. To maintain a hierarchy between the minimum for mass 
difference and the minimum in total mass 
sum during a minimization procedure, we set $\lambda=\min(J_{ij})/\max(P_{ij})$. This choice is based on empirical studies as in the case of choosing hyperparameters in conventional ML  algorithms. Finally, we swap $A$ and $B$ if the number of particles assigned to $A$ is less than the number of particles clustered into $B$. We maintain the ordering between 
numbers of constituent particles in $A$ and $B$ over all events.

\begin{figure}[t]
\centering
\includegraphics[width=0.48\textwidth]{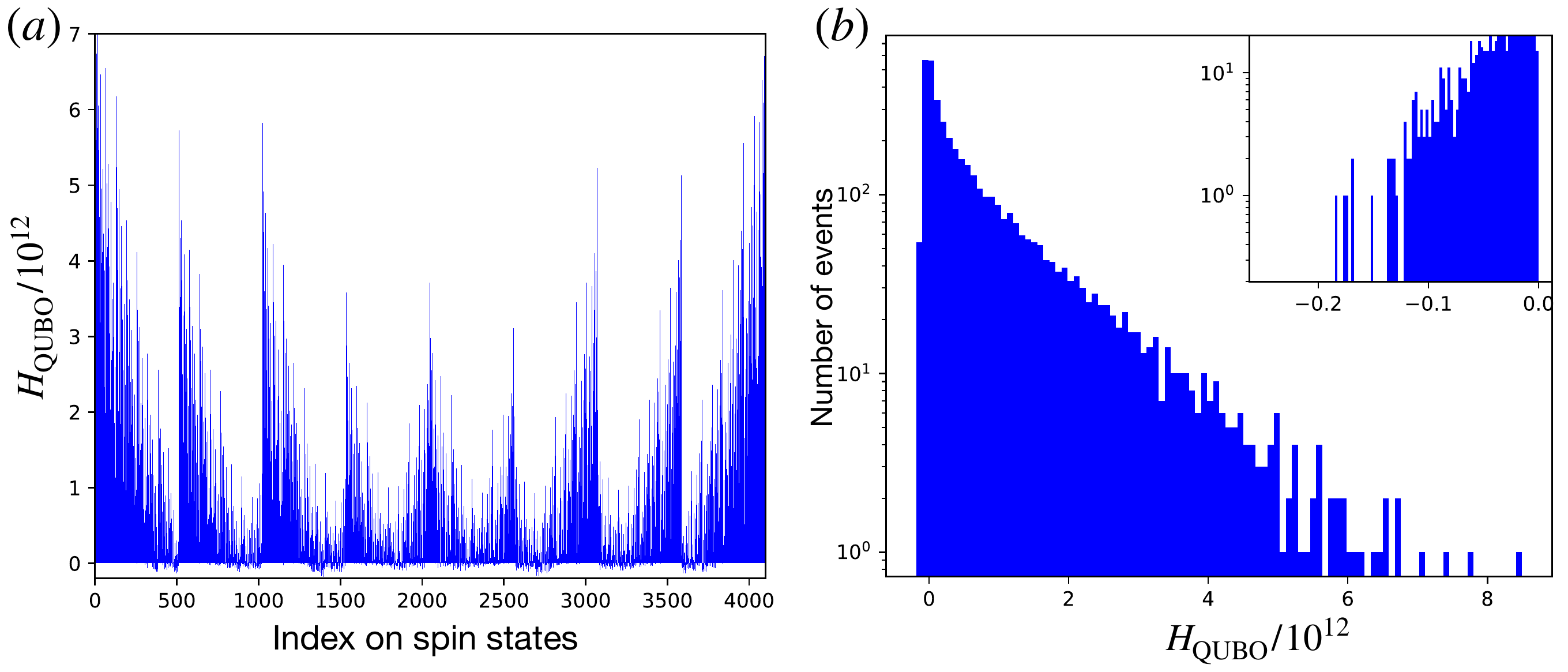}
\caption{We choose an event from detector level MC samples of EEq.q.\,\eqref{eq:fourtop} to calculate (a) energy spectrum of $H_\QUBO$ with increasing indices of spin states, (b) histogram of energy spectrum of $H_\QUBO$ for all possible $2^{12}(=4096)$ spin states.
\label{fig:energy_spectrum}
}
\end{figure}

In order to demonstrate the performance of our QUBO algorithm, we take three examples: (1) Top quark pair production, (2) Higgs and $Z$ boson production and (3) four top-quark production
via the pair of color octet $\tilde o$ scalar where each scalar decays into a top-quark pair\,\cite{Choi:2008ub}. Here we take the mass of $\tilde o$ as $600\,\rm{GeV}$ for a benchmark.
All these particles decay hadronically;
\begin{subequations}
\label{processes}
\begin{align}
&pp\to t,\bar t \to \{j_1,j_2,j_3,j_4,j_5,j_6\}, \label{eq:ttbar}\\
&pp\to H, Z \to   \{j_1,j_2,j_3,j_4,j_5,j_6\},  \label{eq:HZ}\\
&pp\to \tilde o, \tilde o^* \to t,\bar t, t, \bar t \to \{ j_1,j_2,j_3,\cdots,j_{11}, j_{12}\}. \label{eq:fourtop}
\end{align}
\end{subequations}
Here $j_i$ is a reconstructed jet as we deal with fully hadronic processes. 
To prepare data for above processes at the LHC@13TeV, we use the standard chain of Monte Carlo simulations,  {\sc MadGraph5}, {\sc Pythia8} and {\sc Delphes3} with {\sc Fastjet}\,\cite{Alwall:2014hca, Sjostrand:2014zea, deFavereau:2013fsa, Cacciari:2011ma}. As we focus on testing the feasibility of our QUBO algorithm, we apply it to signal processes with MPI and ISR/FSR processes turned off. Jets are reconstructed through anti-$k_T$ algorithm with a jet radius $R=0.4$. Basic cuts of $p_T > 25\,\rm{GeV}$ and rapidity $|\eta| < 2.5$ are applied to reconstructed jets.
$H_\QUBO$ in Eq.\,\eqref{eq:QUBO_lambda} is calculated with Monte Carlo data for a given spin state $\{s_i\}$. In Fig.\,\ref{fig:energy_spectrum} we show (a) the energy spectrum of $H_\QUBO$  and (b) histogram of energy spectrum of $H_\QUBO$ with an event from a four top-quark production process as in Eq.\,\eqref{eq:fourtop}. 
For the ordering of spin states in Fig.\,\ref{fig:energy_spectrum}(a), we increase a spin state by flipping a spin in an increasing order based on a binary digit. For example with four spins, the spin order $(\uparrow\uparrow\uparrow\uparrow\rightarrow\uparrow\uparrow\uparrow\downarrow\rightarrow\uparrow\uparrow\downarrow\uparrow\rightarrow \uparrow\uparrow\downarrow\downarrow\rightarrow\cdots)$   corresponds to the index as $(0 \to 1 \to 2\to 3\to 4\to \cdots)$.
One can try a conventional procedure called a simulated annealing to find a global minimum in $H_\QUBO$ distribution of Eq.\,\eqref{eq:QUBO_lambda}\,\cite{doi:10.1126/science.220.4598.671}. 
Simulated annealing uses a thermodynamic probability to find a global ground state. 
It starts with an initial temperature $T_0$ and gradually decreases temperature $T$ to zero degree
at each annealing step. In each step, this algorithm checks whether flipping a spin is beneficial to get a global minimum. 
If the energy with flipped spin is lower than the initial energy, it takes the flipped spin configuration. 
If not, the spin will be flipped according to the probability of Boltzmann factor, $e^{-(E_{n+1}-E_{n})/{k_B\,T}}$. But when the structure of an energy spectrum with a spin state is complicated,
it will have a local minimum problem. In our case, the energy spectrum can be extremely complicated as shown in Fig.\,\ref{fig:energy_spectrum}.  In Fig.\,\ref{fig:energy_spectrum}(a),  the energy structure similar to a dense pine tree park neutralizes simulated annealing, as sudden drops and rises disable the attempt of spin flipping procedures.  On top of this local minimum problem, the population near a global minimum is sparse as we observe in Fig.\,\ref{fig:energy_spectrum}(b). 
Thus we choose to take a quantum advantage to find a global minimum for a complicated energy distribution. 
\begin{figure*}[t!]
\centering
\includegraphics[width=0.45\textwidth]{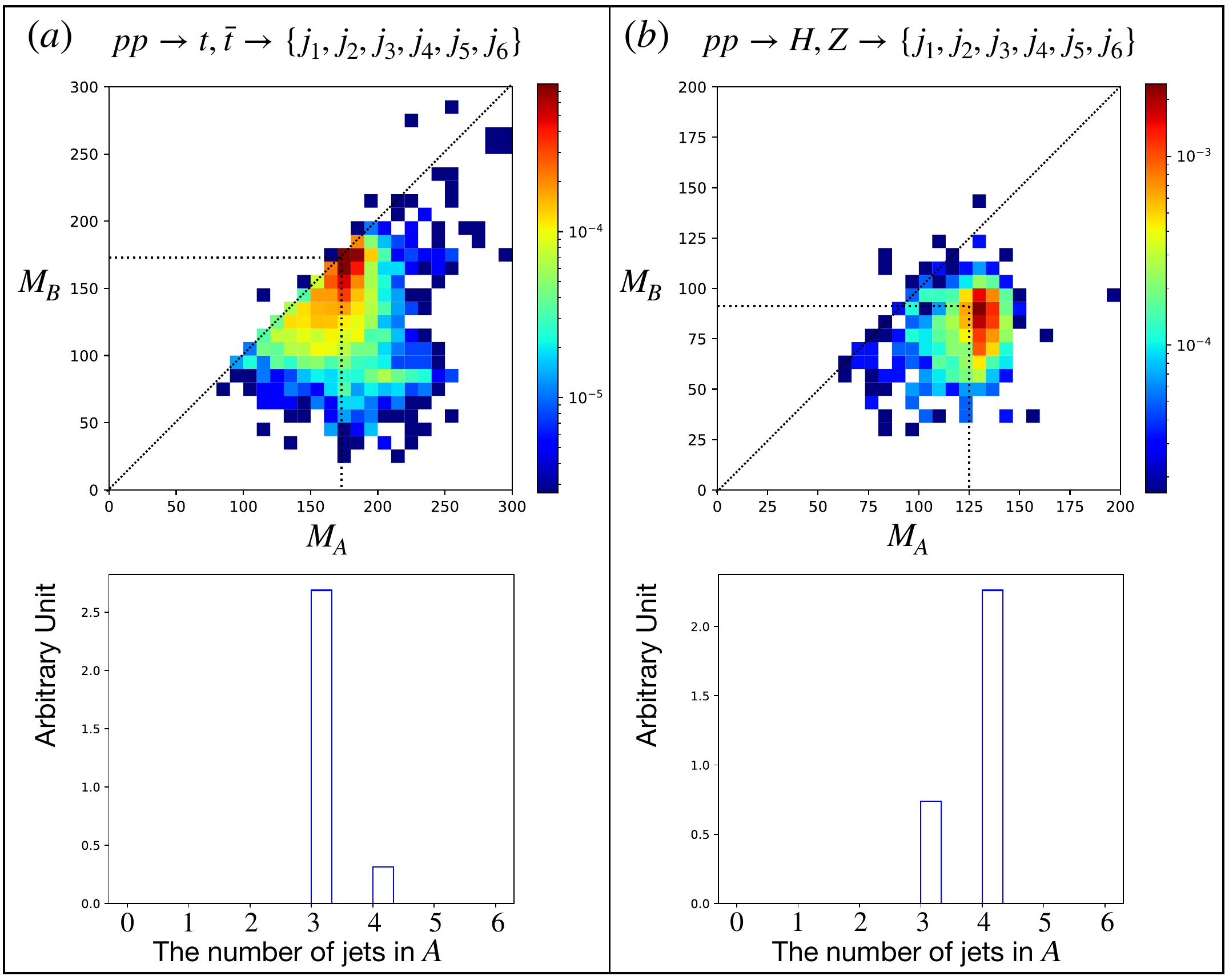}
\includegraphics[width=0.45\textwidth]{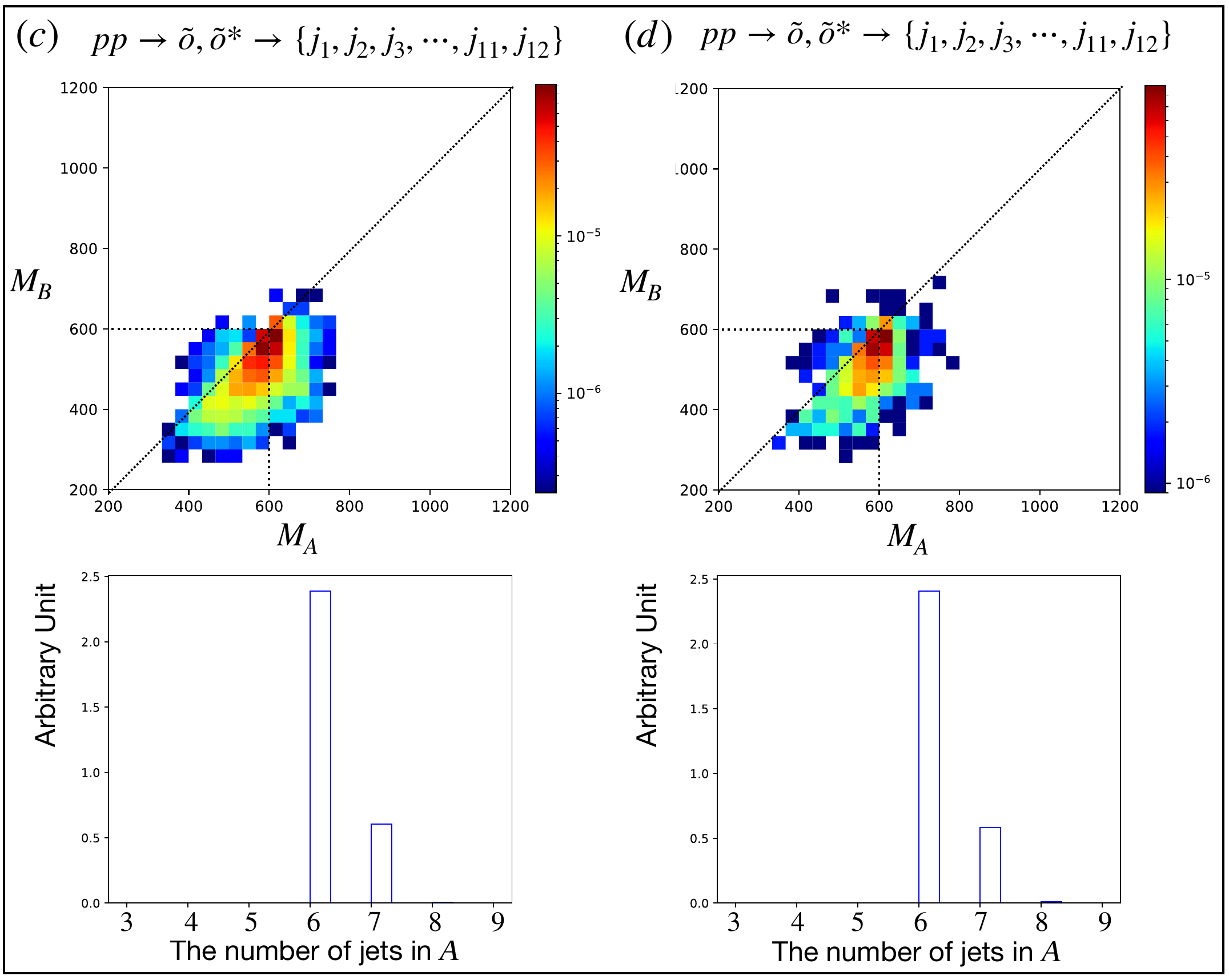}
\caption{\textbf{Common:} $H_\QUBO$ is calculated with Monte Carlo samples. To find a global minimum of $H_\QUBO$, we use a brute force scanning in (a)-(c), and quantum annealer in (d). For (d), we randomly choose 1000 events from MC samples we used in (c). \\
\textbf{Top:} Normalized density histogram for reconstructed mass $M_A$ and $M_B$ 
\textbf{Bottom:} \label{fig:result} The number of jets clustered into $A$.
}
\end{figure*}

\noindent{\bf Quantum advantage.} 
Quantum annealing (QA) is optimized to handle problems in a QUBO form.
It uses the adiabatic theorem to find the ground state of a complicated $H_{\rm QUBO}$ starting from the ground state of a trivial Hamiltonian $H_0$\,\cite{PhysRevE.58.5355};
\beq
H_{\rm QA}(t) = A(t) H_0 + B(t)H_{\rm QUBO},
\label{eq:QA}
\eeq
where $H_0 = \sum_i \left(s_\perp\right)_i$ with a new spin set $\{s_\perp\}$ which is 
transverse to the spin set $\{s\}$ of $H_{\rm QUBO}$. 
At the beginning of $t=0$,  $H_{\rm QA}(0) = A(0) H_0$ as
$A \neq 0$ and $B=0$. Thus the ground state of $H_{\rm QA}(0)$ is the same as the ground state of $H_0$. 
By adiabatically decreasing $A$ to 0 but increasing $B$ with a time $t$, the ground state of $H_0$ can be transmitted to the ground state of $H_{\rm QUBO}$ via $H_{\rm QA}$.
To realize QA process of Eq.\,(\ref{eq:QA}), we use a commercial D-Wave Advantage{\texttrademark} which has $5000+$ available spins (=qubits)\,\cite{Dwave.Avantage.Overview}.

\begin{figure}[t]
\centering
\includegraphics[width=0.47\textwidth]{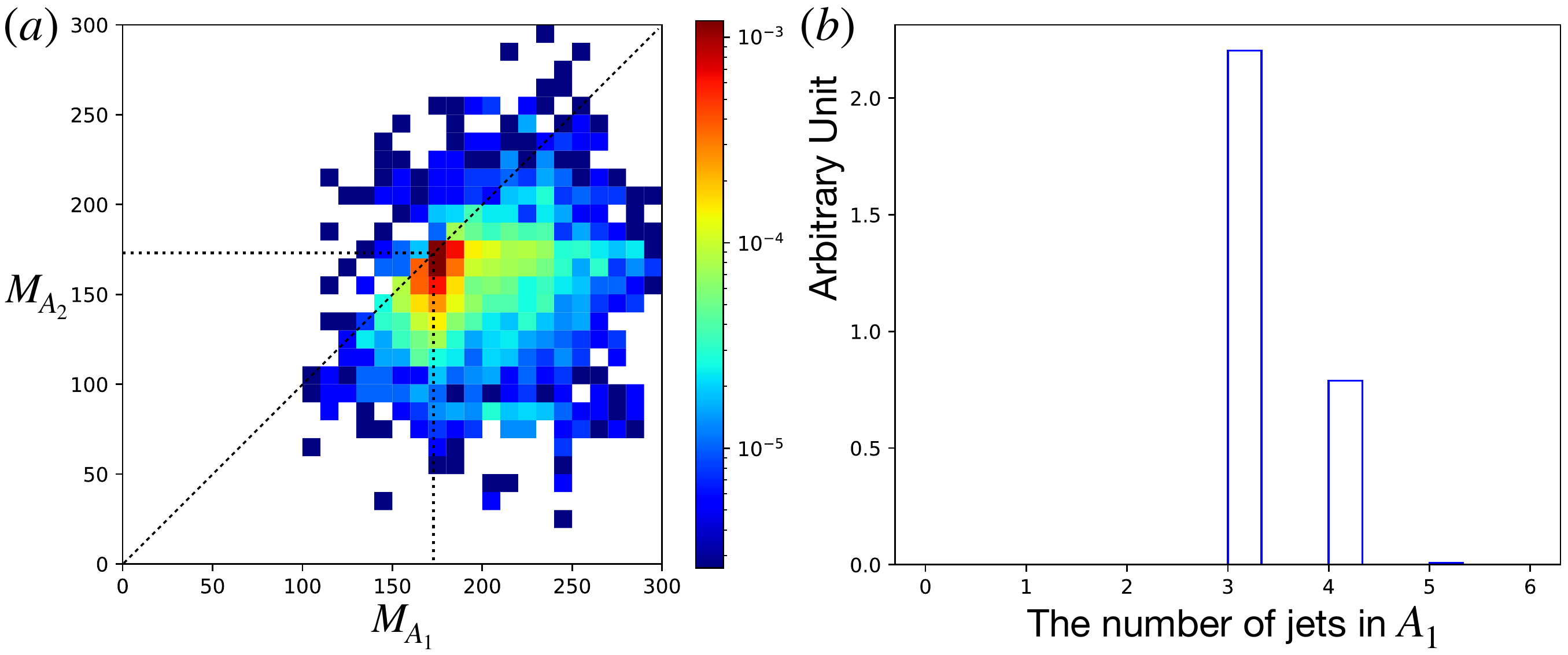}
\caption{Sequential application of $H_\QUBO$ to zoom in on the structure of $A$ in Eq.\,\eqref{eq:fourtop}.  After 12 jets are divided into two groups $A$ and $B$, $H_\QUBO$ further investigates the structure of $A$. 
It identifies that $A \to t \bar t$ by measuring (a) masses and (b) the number of decaying particles of $A_1$, $A_2$ for $A\to A_1, A_2$.
\label{fig:substructure}
}
\end{figure}

Most of time spent by a QA procedure is dedicated to a preparation step, while required time for an annealing process is independent on the size of inputs.
In our case with Eq.\,(\ref{eq:QUBO}),  preparation time 
$T_{\rm QUBO}$ is of $\mathcal{O}(n^2)$.
Compared to the processing time of $\mathcal{O}(2^n)$ with the simplest but a robust brute-force scanning algorithm with a classical computer, a quantum annealer can have an enormous advantage in  
the computational complexity as
\beq
T_{\rm QUBO}(n)\sim \mathcal{O}(n^2) \ll \mathcal{O}(2^n),
\eeq
\begin{table}[t]
\begin{tabular}{|c|c|c|c|}
\hline
\multirow{2}{*}{Process}&$ pp\rightarrow t\bar{t} $&$p p\rightarrow H Z$&$pp\rightarrow \tilde o \tilde o^*$\\
 & Eq.\,\eqref{eq:ttbar} & Eq.\,\eqref{eq:HZ} & Eq.\,\eqref{eq:fourtop} \\
\hline
Success rate&100\% &100\% &93\% \\
\hline
\end{tabular}
\caption{Success rate in finding a global minimum of $H_\QUBO$ using D-Wave Advantage{\texttrademark}. 
\label{dwave_result}}
\end{table}
In Table\,\ref{dwave_result}, we illustrate the performance of a quantum annealer in finding a global minimum. 
Monte Carlo samples for $H_\QUBO$ are generated as in the previous section.
As we notice, current quantum annealer achieves a good performance to find a global minimum for complicated energy distributions which is not possible with simulated annealing. 
By assigning jets into either $A$ or $B$, we can reconstruct the four-momenta of $A$ and $B$ to identify their properties as in Fig.\,\ref{fig:result}.
Reconstructed mass $M_A$ and $M_B$ with $H_\QUBO$ algorithm spots the true mass point (Top panel in  Fig.\,\ref{fig:result}). The most populated number of clustered jets in $A$ is equal to the true number of decayed particles from $A$ (Bottom panel in  Fig.\,\ref{fig:result}) for a hadronically decaying top quark in Eq.\,\eqref{eq:ttbar}, a higgs decaying into four jets via $W^\pm$ bosons in Eq.\,\eqref{eq:HZ} 
and a color octet scalar $\tilde o$ which decays into a top-quark pair, resulting in six jets as in Eq.\,\eqref{eq:fourtop}.  
We can apply $H_\QUBO$ sequentially to find the substructures of $A$ and $B$;
\bea
H_\QUBO^{(A)}&=&\sum_{ij=1}^\ell J'^{\alpha}_{ij}s^\alpha_is^\alpha_j+\sum_{i=1}^\ell h'^\alpha_i s^\alpha_i,\\
H_\QUBO^{(B)}&=&\sum_{ij=1}^m J'^\beta_{ij}s^\beta_is^\beta_j+\sum_{i=1}^m h'^\beta_i s^\beta_i,
\eea
where $\{s_i^\alpha\}$ is a spin set for particles clustered into $A$ and $\{s_i^\beta\}$ is the one for particles assigned to $B$ after minimizing an original $H_\QUBO$. Here $\ell$ and $m$ vary in an event by event basis, only need to satisfy $\ell+ m=n$. We get additional constraints for the number of intermediate particles from the decay of each of $A$ and $B$ as $A\to A_1, A_2$ and $B\to B_1, B_2$. 
In Fig.\,\ref{fig:substructure}, we present the result of above sequential application to the most complicated process of Eq.\,\eqref{eq:fourtop}.
Sequential QA reveals the structure of an event-topology behind 12 jets as;
\bea
pp \to &&\tilde o \tilde o^*, \left(\tilde o \to 6j \right), \left(\tilde o^* \to 6j \right) \textrm{ with} \nonumber \\
 &&\tilde o \to t \bar t, \left( t \to 3j\right), \left( \bar t \to 3j\right),  \nonumber \\
 && \tilde o^* \to t \bar t, \left( t \to 3j\right), \left( \bar t \to 3j\right),  \nonumber 
\eea
by measuring masses and the number of constituent jets of $A_1$ and $A_2$ as well as $B_1$ and $B_2$. 

\begin{figure}[t]
\centering
\includegraphics[width=0.47\textwidth]{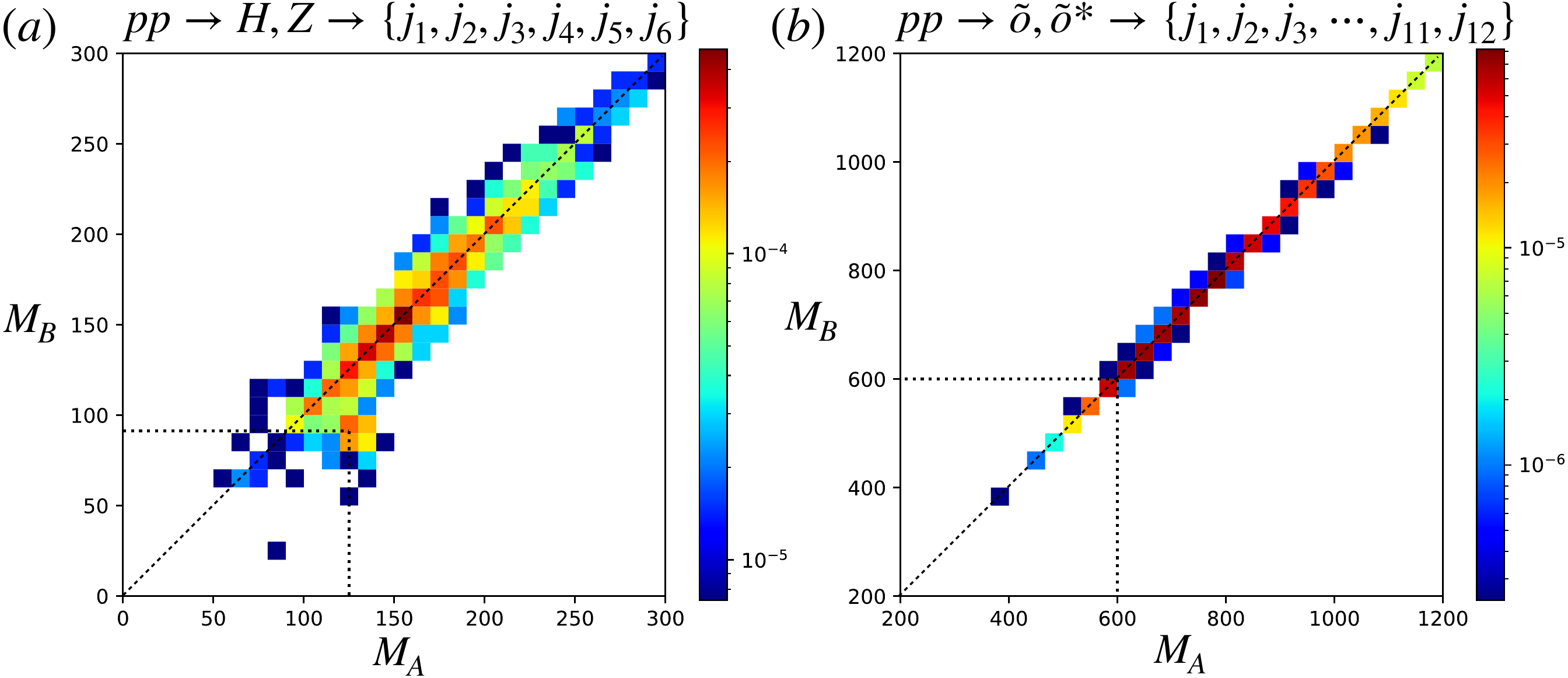}
\caption{Results with same MC samples in Fig.\,\ref{fig:result} but using $H_\QUBO$ as in Eq.\,\eqref{eq:QUBO} without a constraint term. QUBO algorithm in this case fails in spotting right mass spectrum for (a) asymmetric case, and (b) highly smeared and off-shell case.
\label{fig:withoutLambda}
}
\end{figure}
Before closing this section, we explain the effect of a constraint term $\lambda(P_1^2+P_2^2)$ in Eq.\,\eqref{eq:QUBO_lambda} by showing results only with minimizing differences between $M_A$ and $M_B$ without the constraint term in Fig.\,\ref{fig:withoutLambda}. As we expect, $H_\QUBO$ in Eq.\,\eqref{eq:QUBO} focuses on minimizing the mass 
difference between $A$ and $B$ which is inadequate in handling situations including asymmetric processes like $\left(pp\to H,Z\right)$ in Eq.\,\eqref{eq:HZ}, particles with a large decay width, and experimental defects including smearing effects mostly for multi-jet productions as in Eq.\,\eqref{eq:fourtop}.
\begin{table}[h]
\centering
\begin{tabular}{|c|c|c|c|c|}
\hline
\multicolumn{2}{|c|}{\multirow{2}{*}{Process}} & $ pp\rightarrow t\bar{t} $&$p p\rightarrow H Z$&$pp\rightarrow \tilde o \tilde o^*$\\
 \multicolumn{2}{|l|}{}  & Eq.\,\eqref{eq:ttbar} & Eq.\,\eqref{eq:HZ} & Eq.\,\eqref{eq:fourtop} \\
\hline
\multirow{2}{*}{Algorithm}& QUBO& 47.3\%& 89.5\%& 17.4\%\\
 \cline{2-5}
 & Hemishere& 33.6\%& 86.2\%&7.2\%\\
\hline
\end{tabular}
\caption{Matching accuracy for the reconstructed momenta of particles $A$ and $B$ using a clustering algorithm to an actual momenta of $A$ and $B$  (parton level analysis but with same basic cuts as in previous Monte Carlo samples).
\label{tab:matching_result}}
\end{table}

We close this section by comparing our algorithm and an existing one. In fact, the subject of identifying event-topology has not gained much attention as the LHC studies were focused more on optimizing
 discovery chances of theoretically well motivated models, mostly supersymmetric ones where relevant event topologies are manifest\footnote{Identifying an event topology in missing energy channel was introduced in\,\cite{Cho:2012er}}.
 If we narrow down to a clustering problem in separating decay chains, there is a hemisphere algorithm that was designed to assign visible particles correctly according to a presumed event-topology\,\cite{CMS:2007sch, Matsumoto:2006ws}. To compare the performance between QUBO and hemisphere method, we use a parton level Monte Carlo samples. In Tab.\,\ref{tab:matching_result}, we present a matching accuracy by 
counting events where all particles are correctly clustered.
 As we observe, the matching accuracy is not high unlike the performance in finding a global minimum of $H_\QUBO$ in Tab.\,\ref{dwave_result}.

\begin{figure}[t!]
\centering
\includegraphics[width=0.4\textwidth]{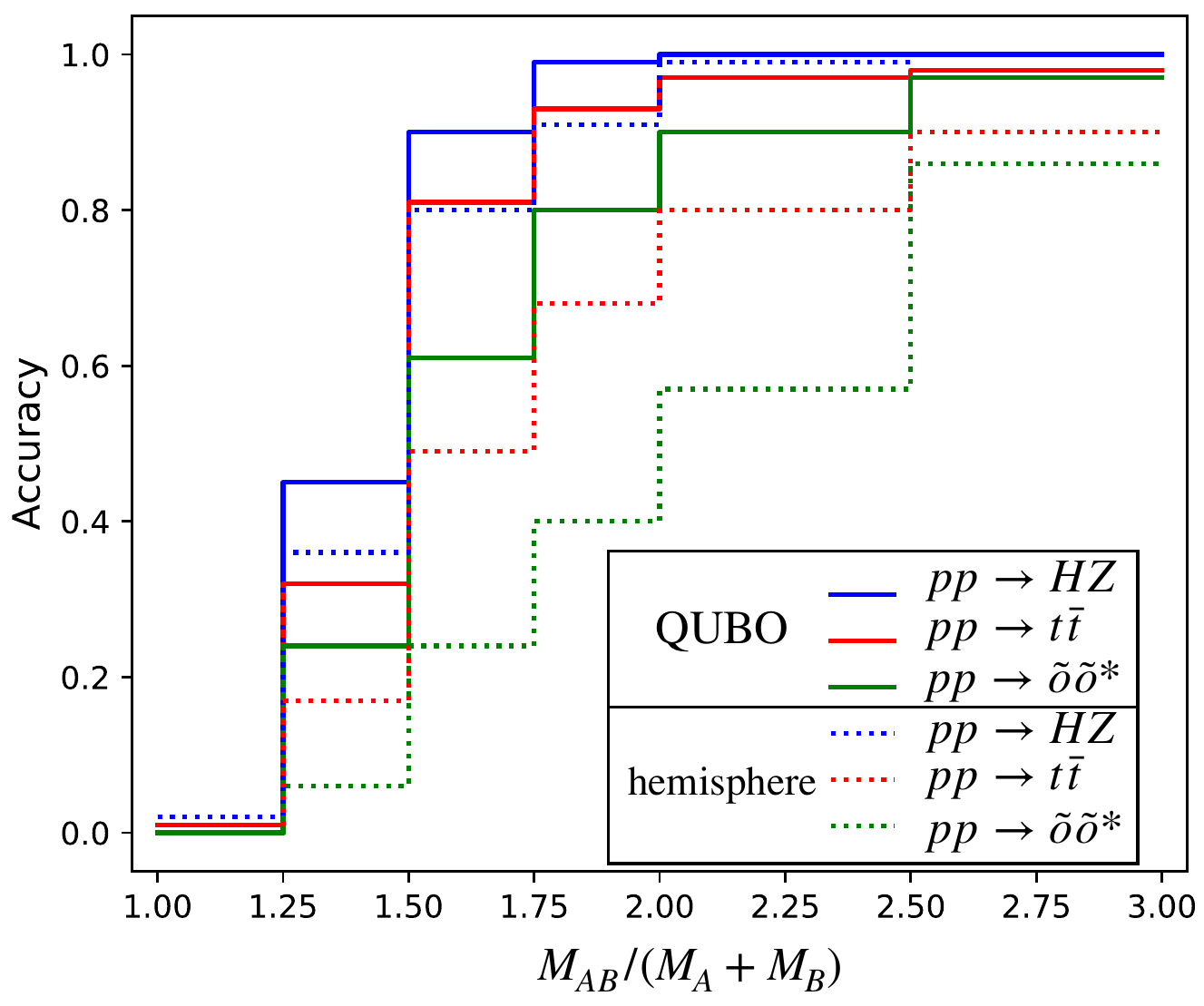}
\caption{Matching accuracy same as in Tab.\,\ref{tab:matching_result} in terms of their boost factor $M_{AB}/{(M_A+M_B)}$. Here $M_{AB}$ is an invariant mass of $A$ and $B$. 
\label{fig:accuracy}
}
\end{figure}
In Fig.\,\ref{fig:accuracy}, we trace the matching accuracy with a boost factor of $A$ and $B$ to understand this gap
\footnote{The Lorentz boost factor $\gamma$ for a particle A is $E_A /M_A = M_{AB}/(2M_A)$. To deal with the case of $M_A \neq M_B$, we take the approximate average of $\gamma_A$ and $\gamma_B$ as $M_{AB}/(M_A+M_B)$.}. 
When $A$ and $B$ are at rest as $M_{AB} / (M_A + M_B) = 1$, there is no kinematic structure nor any hint to assign particles into their ``mother" particle. 
As $A$ and $B$ get a boost, particles from them develop a pencil-like structure for clustering algorithms to utilize. 
We see this tendency in Fig.\,\ref{fig:accuracy}. 
Most of top and anti-top quarks are not boosted as they are produced closed to threshold while $H$ and $Z$ are moderately boosted beyond their threshold.
On top of this, a huge difference between $m_W$ and $m_{j_b}$ in the decay of a top-quark $(t\to b,W^+)$ weakens a pencil structure compared to a simple kinematic structure in the decays of $Z$ and $H$. These explain degraded performances of clustering algorithms in a $t\bar t$ process compared to the case of $HZ$. 
One interesting point is that our QUBO algorithm is a ``seedless" method. A hemisphere method takes a particle with the highest momentum as a seed, and requires the second seed to separate a phase space into two. Thus a hemisphere method is weaker than QUBO in the region of low boost and in the case of a complicated decay structure.
Any method based on a seed will become powerless with increasing number of particles where particles share a phase space uniformly as in the process of Eq.\,\eqref{eq:fourtop}. Here we emphasize that our QUBO algorithm is constructed from the very minimal assumption without relying on the kinematic structure. This provides an advantage to QUBO compared to previous ones which are based on the geometry of a preassumed phase space. 

\noindent{\bf Conclusion.} In this letter, we illustrate how a bottom-up approach with data from high energy colliders can be established via a quantum algorithm. 
To have the full advantage of our method, a quantum computer is necessary in finding the global minimum of a complex energy distribution.  
The technologies of a quantum computer are on the verge of quantum supremacy\,\cite{QM1, QM2}. Thus as a theorist, it is our duty to formulate problems into a right form to have the full benefits of a quantum computer in decoding the fundamental laws of physics. As a TeV-scale high energy collider can reach the moment of $\mathcal{O}(10^{-12})$ seconds after big-bang, the realm of ``quantum universe", Nature will reveal its secrets when we face it with quantum technologies. 

\section{acknowledgments}
This work is supported in part by KIAS Individual Grant No.\,QP078301 (MK)
via the Quantum Universe Center at Korea Institute for Advanced Study, 
by KIAS Individual Grants under Grant No.\,PG021403 (PK), 
by Basic Science Research Program through National Research Foundation of Korea (NRF) 
Research Grant NRF-2019R1A2C3005009 (PK, JhP) and by NRF-2021R1A2C4002551 (MP). 
\newline
\appendix*
\section{Quantum annealer}
\begin{figure}[h]
\centering 
\includegraphics[width=0.3\textwidth]{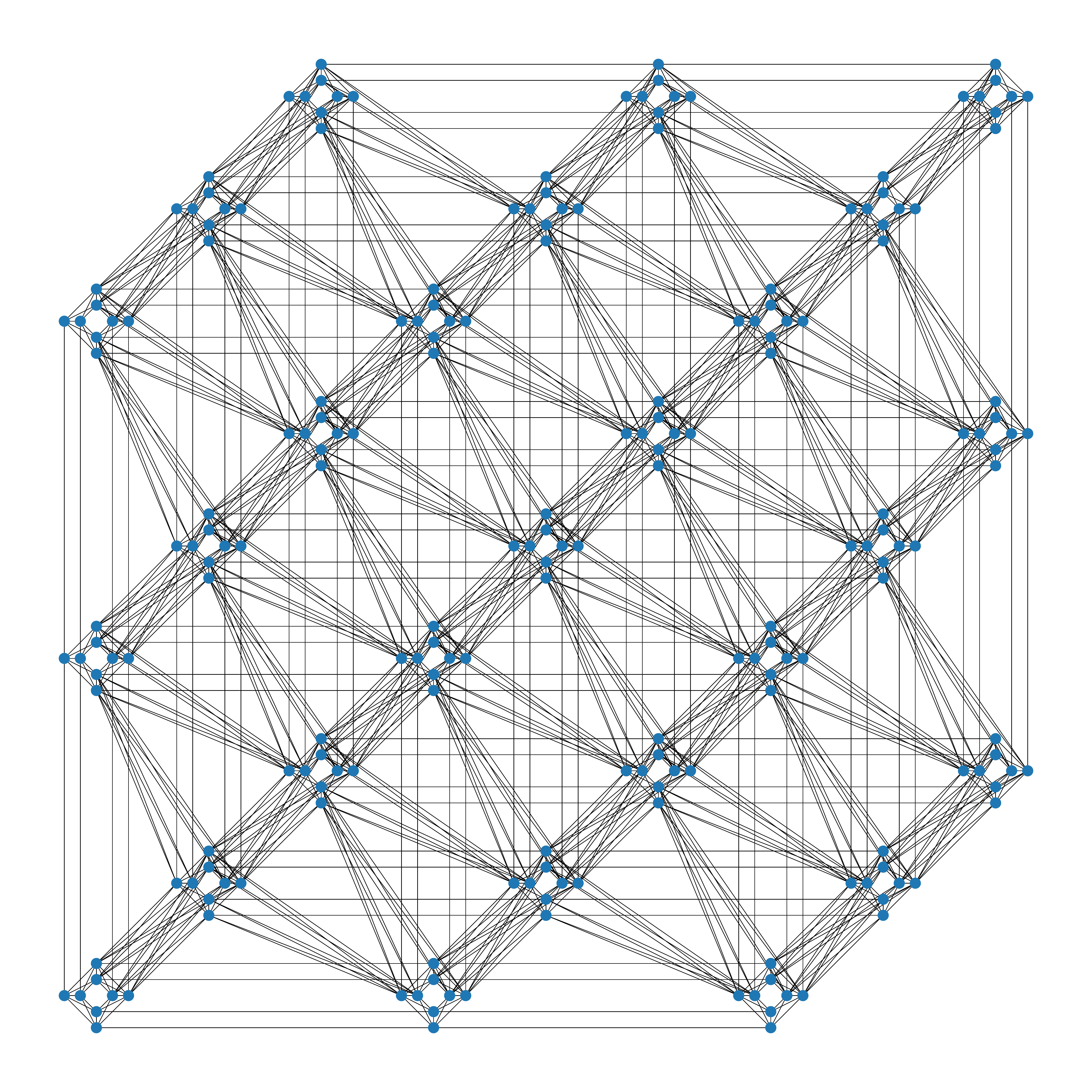}
\caption{A Pegasus graph with 27 unit cells. 
\label{fig:dwave_graph_pegasus}}
\end{figure}
We use Amazon Web Services (AWS) to access D-wave Advantage{\texttrademark} 
(Advantage), a quantum annealer by D-wave company.
Advantage has 5000+ qubits, 
connected to each other with at least 35000 couplers 
which are way fewer than the number of all possible 
pairs ${5000+ \choose 2}$.
Thus the network among qubits is not fully connected. To squeeze performance in this limited situation, Advantage has a connectivity structure among qubits, called as 
Pegasus graph\,\cite{boothby2020nextgeneration}. 
This graph consists of unit cells each containing 8 qubits.

\begin{figure}
\centering 
\includegraphics[width=0.48\textwidth]{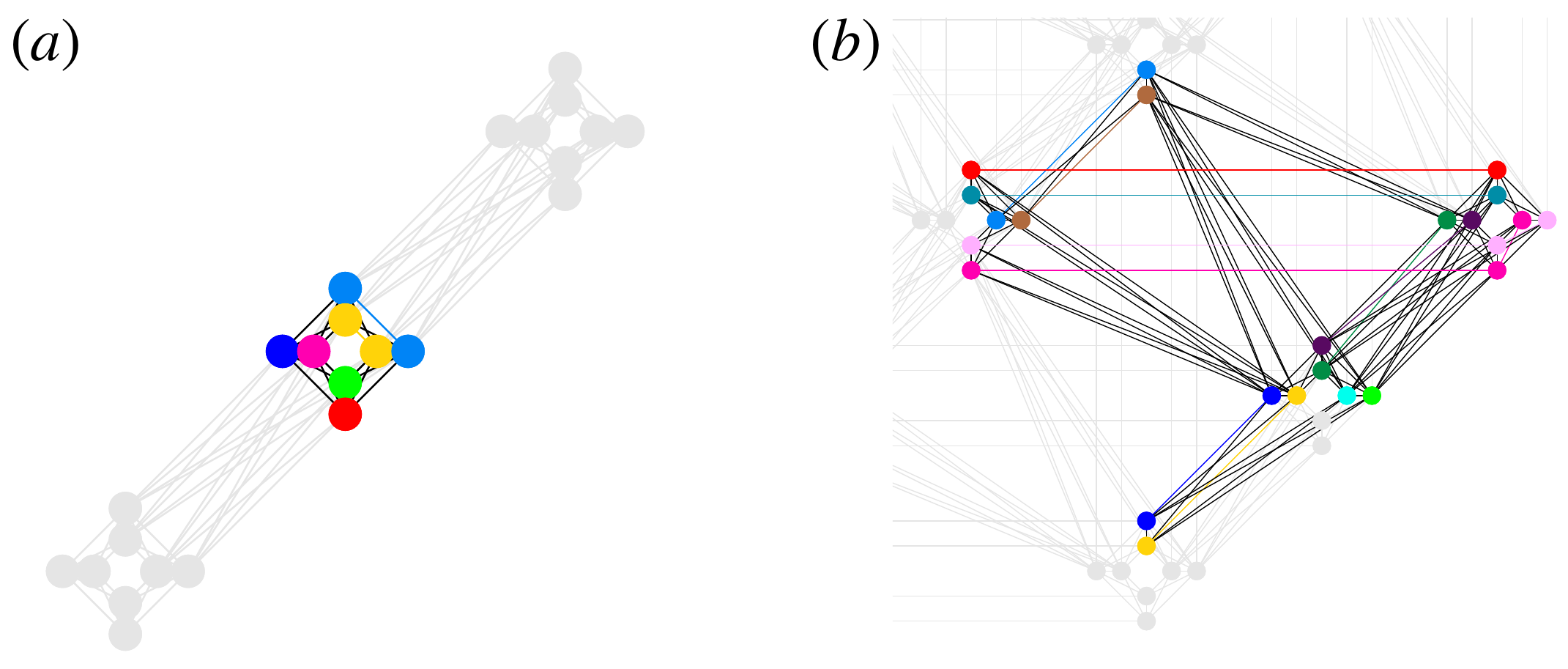}
\caption{A fully connected chain in Pegasus graph requires (a) 8 qubits for 6 visible particles from processes in\,\crefrange{eq:ttbar}{eq:HZ} , and
(b) 24 qubits for 12 visible particles in Eq.\,\eqref{eq:fourtop}.
\label{fig:spinchain}
}
\end{figure}

Fig.\,\ref{fig:dwave_graph_pegasus} shows a partial sample of Pegasus graph.
The network structure of Pegasus is fixed with varying strength in couplers between connected qubits. 
To embed various QUBO problems into this fixed and non-fully connected network of qubits, Advantage uses a minor-embedding with chains of qubits\,\cite{cai2014practical}. 
For example, one needs to have a qubit which is connected to both $q_i$ and $q_j$, but can not find this one in a given network. 
In this case, it would be easier to find a chain (a set of spins, connected with each other like a chain), 
$\{q_1, q_2, \cdots, q_n\}$ where $q_1$ is connected to $q_i$ and $q_n$ has a connection to $q_j$.
The connection strengths in a chain should be chosen properly so that qubits in a chain are treated as a single qubit during solving a QUBO problem. 
Advantage provides an automatic embedding solution with adjustable chain strength parameters\,\cite{Dwavedocumentation}. 
Thus the required number of qubits is larger than the number of inputs in a given QUBO problem. In our case, we need 8 qubits to solve our QUBO problems with 6 visible particles for processes in\,\crefrange{eq:ttbar}{eq:HZ} 
and 24 qubits for 12 particles in a process of Eq.\,\eqref{eq:fourtop} as in Fig.\,\ref{fig:spinchain}.

As a quantum annealer 
has various systematic errors especially in the status of qubits during operations, a performance of Advantage will drop with increasing number of required qubits. 
The required number of qubits in Eq.\,\eqref{eq:fourtop} is three times larger compared to the one in\,\crefrange{eq:ttbar}{eq:HZ} while the number of input visible particles only increases by the factor of two.
To reduce errors, a method which is called the spin reversal transform was introduced\,\cite{pelofske2019optimizing}.
The spin reversal transform flips the sign of selected coefficients of QUBO problem which 
Advantage minimizes. 
The spin reversal transform does not change the ground state of QUBO problem, 
so one can average out the biases which come from systematic errors. 
We use default spin chain strength parameter and set 
the number of spin reversal transform to 10 to minimize systematic errors. 
A recent work suggests that one can increase the success rate in QA
through a parallelization procedure which processes several events in a single 
QA operation\,\cite{pelofske2021parallel}.
\bibliography{reference}
\end{document}